\documentclass[pra,preprint,aps,superscriptaddress]{revtex4}

\usepackage[T1]{fontenc}
\usepackage{amsmath}
\usepackage{graphicx}
\usepackage{tabularx}
\usepackage{amssymb}
\usepackage{hyperref}
\usepackage[usenames]{color}
\definecolor{rltred}{rgb}{0.75,0,0}
\definecolor{rltgreen}{rgb}{0,0.5,0}
\hypersetup{colorlinks,linkcolor=rltred,citecolor=rltgreen}

\begin{document}

\title{A grid-based Ehrenfest model to study electron-nuclear processes}

\author{Bo Y. Chang}
\affiliation{School of Chemistry, Seoul National University, Seoul 08826, Republic of Korea}

\author{Seokmin Shin}
\affiliation{School of Chemistry, Seoul National University, Seoul 08826, Republic of Korea}
\email{sshin@snu.ac.kr}

\author{Vladimir S. Malinovsky}
\affiliation{U. S. Army Research Laboratory, Adelphi, Maryland 20783, USA}

\author{Ignacio R. Sola}
\affiliation{Departamento de Qu\'imica F\'isica, Universidad Complutense, 28040 Madrid, Spain}
\email{isola@quim.ucm.es}

\begin{abstract}
The two-dimensional electron-nuclear Schr\"odinger equation
using soft-core Coulomb potentials has been a cornerstone 
for modeling and predicting the behavior of one-active-electron diatomic
molecules, particularly for processes where both bound and continuum states
are important. The model, however, is computationally expensive to extend 
to more electron or nuclear coordinates. Here we propose to use the
Ehrenfest approach to treat the nuclear motion, while the electronic
motion is still solved by quantum propagation on a grid.
In this work we present results for a one-dimensional treatment of H$_2^+$,
where the quantum and semi-classical dynamics can be directly
compared, showing remarkably good agreement for a variety of situations.
The advantage of the Ehrenfest approach is that it can be
easily extended to treat as many nuclear degrees of freedom as needed.
\end{abstract}

\maketitle

\section{Introduction}


Full control over chemical reactions and molecular properties
requires the manipulation of electronic and nuclear
degrees of freedom (dofs). When the system is in the ground electronic
state, the energy and time-scales associated to both dofs is often
very different, allowing for a separate study of their dynamics,
in the spirit of the Born-Oppenheimer approximation.
This ceases to be the case when the dynamics occurs in excited
electronic states, where non-adiabatic processes ({\it i.e.} by
definition beyond Born-Oppenheimer) are the usual suspects behind
the fate of most molecular events leading to energy deactivation
or to a specific channel in a chemical reaction
\cite{YarkonyRMP96,WorthARPC04,DomckeYarkonyKoppel}.
Because of computational constraints, most non-adiabatic processes
have been analyzed expanding the wave function on a small subset
of electronic states
\cite{PersicoTCA14,BeckPR00,Meyer2009,Tully1998,YoneharaCR12,RichterJCTC11,TullyJCP12,MaiWIRE18,BenNunJPCA00,ShalashilinJCP09}. 
However, if the energy in the system is large, so is the density
of states, and different types of models must be elaborated to
fully account for electron and nuclear correlations and their
control, particularly in the presence of strong fields.
 
Recently, there has been a surge of studies that aim to analyze
and control electron-nuclear processes\cite{KimJPCA12,BraunJPB14,CalegariSci14,SolaPCCP15,WolterSci16,WangCPL17,AttarSci17,NisoliCR17,AminiPNAS19}
to create novel transient molecular properties in the presence
of strong fields, for instance huge electronic dipoles
\cite{ChangJCP13,ChangJPB15,ChangJPB15b},
to manipulate non-adiabatic transitions
\cite{HofmannCPL01,JGVCPL06,JGVJCP09,FalgeJPCA12}
or to unravel 
the electron-nuclear dynamical features in conical intersections
\cite{HaderJCP17,ArnoldPRL18,CsehiJPCL17}
particularly at so-called light-induced conical intersections
\cite{MoiseyevJPB08,HalaszJPCA12,DemekhinJCP13,HalaszJPCL15},
Our goal is to design a simple model that can provide qualitative predictions
of quantum control in one-active electron systems, treating strong field laser
couplings, non-adiabatic couplings and ionization on equal footing,
and that can be extended to polyatomic molecules. 
One possible avenue is to use the Multiconfiguration Time-Dependent Hartree 
(MCTDH) method\cite{BeckPR00,Meyer2009}, the Ab-Initio Multiple Spawning
(AIMS) method\cite{BenNunJPCA00,ShalashilinJCP09}
or other schemes that incorporate quantum features to the nuclear
motion\cite{PrezhdoTCA06,ZamsteinJCP12,ZamsteinJCP12b,SubotnikJCP10,CurchodJCP13}.  

Since we are interested in few active electrons (one, in this work)
under a strong field that can lead to a large deformation of
the molecule and its charge distribution over large distances, 
the typical orbital
basis that are used to describe the electron density in Quantum Chemistry
are not well developed for this purpose. Hence, most methodologies,
which are based on expanding few electronic states on a basis, will
not perform adequately.
In this work we will follow a different methodology.
We solve the time-dependent Schr\"odinger equation for the electron
on a grid\cite{KosloffARPC94}, incorporating
the vibrational motions in a semi-classical manner, via a mean-field
Ehrenfest approach\cite{Tully1998,TullyJCP12}

In order to compare the results of the Ehrenfest approach to the
full quantum calculation, in this paper we will focus on the simplest
molecule under strong fields, treated in low-dimensionality
by assuming that all particles are aligned and interact via soft-core Coulomb 
potentials. Although implying strong approximations, the quantum 
one-electron plus one-nuclear dimension Hamiltonian has been shown to
give results in qualitative agreement with experiments, and
as a model it has provided invaluable guidance to find and
to understand new processes of molecules in strong fields\cite{SolaAAMO18}.

We will consider three different situations to test the performance
of the Ehrenfest approach: Laser-free dynamics of a coherent
superposition of electronic states, strong-field laser-driven dynamics 
in the ground state, and the dynamics in the excited dissociative state
under a strong static field.

The first case is interesting because it is well known that the 
dynamics in a superposition state cannot be well reproduced by an Ehrenfest
approach, since each quantum wave packet in the superposition will feel a
different potential than the mean-field potential governing the
evolution of the nuclear trajectory. Because of decoherence, however,
the impact of this difference over the averaged results may be much
smaller than anticipated. 
The opposite situation occurs when the dynamics is adiabatic ({\it i.e.}
slow) under a strong field. Then the nuclei experience an average, 
so-called light-induced potential\cite{YuanJCP78,BandraukJCP81,ChangIJQC16}
(LIP), which can be perfectly reproduced by the Ehrenfest method\cite{Bajo2012JPCA}. 
In the second case we will use optical non-resonant
fields where in principle non-adiabatic effects might be important.
The difference between the quantum and semiclassical results will
quantify to a certain extent the importance of non-adiabatic effects.
In the third case we use strong static fields, where the dynamics
is expected to be fully adiabatic (that is, to occur in a single LIP). 
Here, we expect that
the difference between the quantum and semi-classical results measure
to a certain degree the impact of ionization and dissociation, which
cannot be well reproduced in a mean-field theory.

\section{Model}

A simple yet powerful enough model often used to integrate the dynamics of a system 
composed of two nuclei and one electron under a strong field is provided by the
Hamiltonian of the aligned particles moving under the effect of the
soft-core Coulomb potential\cite{JavanainenPRA88,SuPRA91,KulanderPRA96}. 
Within the range of validity of the model one can treat quantum mechanically
both the electron and nuclear motion and study strong field effects beyond 
the Born-Oppenheimer approximation.
In this work we will treat the nuclear motion classically under the Ehrenfest
approximation testing how the results compared with those obtained in
a fully quantum $2$ dimensional model, under different conditions,
so that one can caliber the quality of the approximation before working in
models with larger degrees of freedom where the quantum results can not
be directly obtained.
In our semiclassical approximation, both the field $E(t)$ and the bond 
distance $R(t)$ are treated as classical variables that enter as parameters
in the potential. For H$_2^+$, 
\begin{equation}
V_{sc}(z;R(t),E(t)) = -\frac{1}{\sqrt{(z-R(t)/2)^2 + \epsilon^2}} 
	- \frac{1}{\sqrt{ (z+R(t)/2)^2 + \epsilon^2}} + \frac{1}{R(t)} -zE(t)
\end{equation}
The electron motion in the $z$ coordinate obeys the time-dependent
Schr\"odinger equation (in a.u.)
\begin{equation}
i\frac{\partial}{\partial t} \psi(z,t) = -\frac{1}{2} \frac{\partial^2}{\partial z^2}
\psi(z,t) + V_{sc}(z;R(t),E(t)) \psi(z,t)
\end{equation} 
whereas the internuclear motion follows Newton equations with the
mean-field electronic potential in the Hellmann-Feynman approximation
\cite{Tully1998}.
\begin{equation}
\frac{d^2}{dt^2}R(t) = -\frac{1}{M} \langle \psi(z,t)  \left| \frac{\partial}
{\partial z}V_{sc}(z;R(t),E(t)) \right| \psi(z,t) \rangle
\end{equation}
where $M$, the reduced mass of the molecular frame, is approximately 
the mass of the proton.

The initial conditions for the dynamical equations can be obtained under
different procedures. 
In the fully quantum TDSE for both electronic and nuclear dof 
the initial wave function, $\Psi(z,R,0)$, will typically be a product state of
the energy eigenstate of the electronic Hamiltonian $\psi_j(z;R)$ (or a 
superposition of different electronic states) times a Gaussian nuclear
wave function, $\chi(R)$, centered at different internuclear distances 
(for instance, the equilibrium bond distance). 
We will reproduce 
laser-free or laser-driven dynamics following the examples of
Chang et al.\cite{ChangJPB15b}

In the semiclassical approach the initial internuclear position and momentum
are obtained by a Monte Carlo sampling from the Wigner distribution of 
$\chi(R)$, $\rho_W(R,P)$.
For each trajectory $k$, one obtains a different set of phase space
coordinates $(R_k(0),P_k(0))$.
However, we sometimes change the distribution to test the sensitivity
of the scheme to the initial nuclear coordinates. In addition, since the 
$\psi_j(z;R)$ are obtained discretized on a grid, by the Fourier grid 
Hamiltonian method\cite{MarstonJCP89}, while the phase space points are
continuous, we choose the electronic wave function $\psi_j(z;\tilde{R}_k)$
at the grid point $\tilde{R}_k$ closer to $R_k$. 
In order to compare the results obtained by the semiclassical method with those
obtained solving the TDSE for the coupled electron-nuclear motion, one
needs to average the results of different trajectories. Eq.(2) is
solved using the split-operator method\cite{FeitJComputP82} while
Eq.(3) is solved using the forth-order Runge-Kutta method.

\section{Results}

%

\subsection{Field-free Dynamics}
\label{A}

A stern test for a mean-field theory to pass is to compare
the results of its dynamics to the quantum case when the initial
wave function is a superposition of different electronic states, in the
absence of a field that couples ({\it i.e.} mixes) the potentials.
Although no single Ehrenfest trajectory can reproduce the average
observables in this situation, the ensemble average of the trajectories
may give reasonable results due to the decoherence that the
vibrational motion induces on the electronic dynamics.
Focusing only on the electronic dof, 
the decoherence 
provokes dumping in the oscillation of the average electron position 
$\langle z(t) \rangle$, because the interfering pathways that
create the oscillations between the electronic states become dephased 
due to the different vibrational periods in each state\cite{ChangJPB15}.

We first consider the dynamics starting in a symmetric superposition
state $\psi(z;R(0)) = \frac{1}{\sqrt{2}} \left( \psi_1(z;R(0))) +
\psi_2(z;R(0)) \right)$, where $\psi_j(z;R)$ ($j=1,2$) are the ground/first 
excited electronic states of H$_2^+$ in the soft-core Coulomb potential.
In the full quantum results we use for the initial nuclear wave function
a Gaussian wave packet $\chi(R)$ with $0.5$ $a_0$ width, centered at 
$R_0 = 2$ $a_0$, that approximately represents the ground state of the parent
$H_2$ molecule before ionization.

\begin{figure}
\includegraphics[width=8cm]{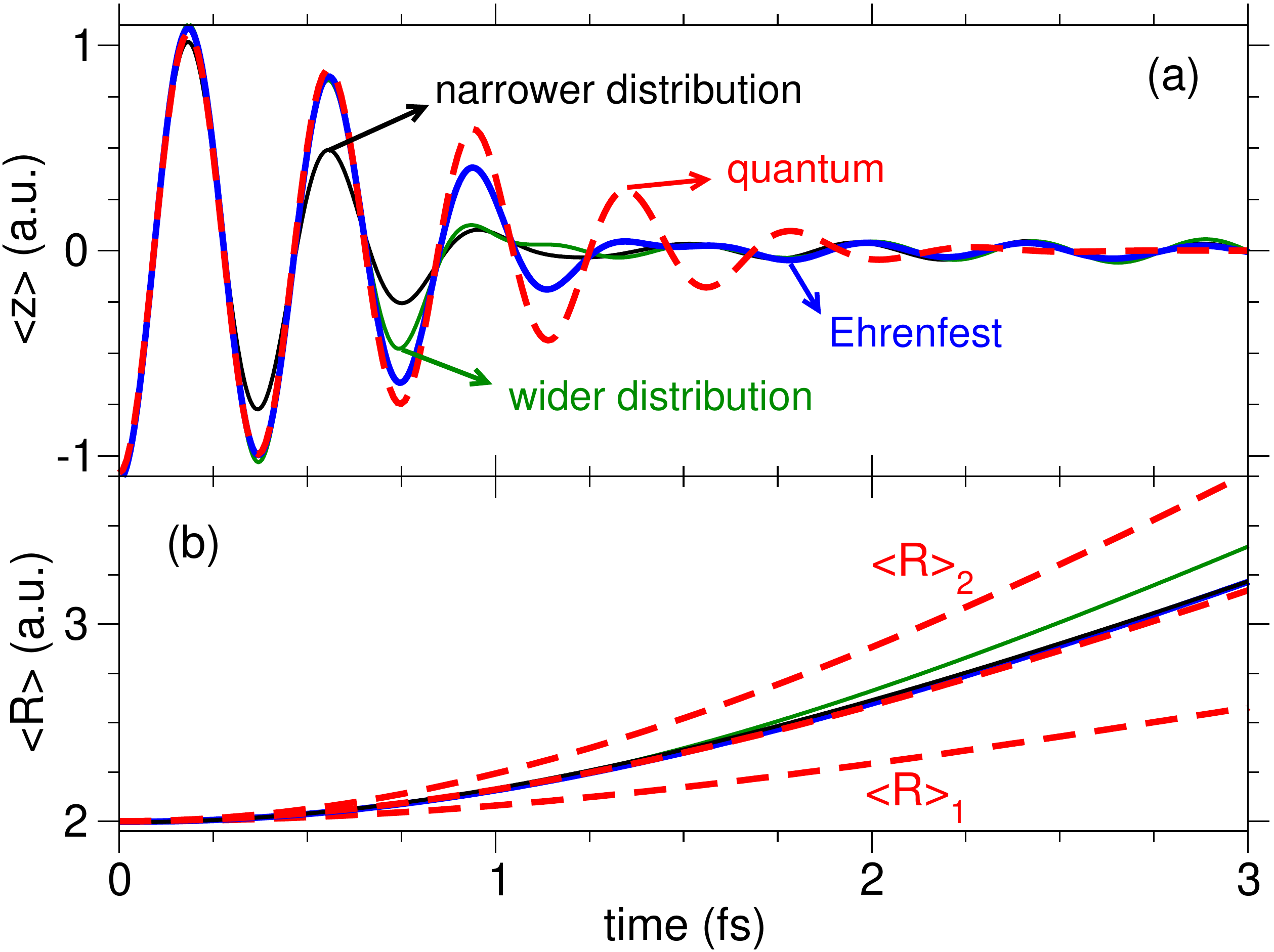}
\caption{Dynamics of the average electron dipole $\langle z\rangle$ (a) 
and internuclear distance $\langle R \rangle$ (b) solved using the TDSE
(red dashed line) and using the Ehrenfest approach from $10000$ trajectories 
sampled from different Wigner distributions: in blue the Wigner corresponding
to the nuclear wave packet and in black and green narrower and wider 
distributions in the position representation. The results are qualitatively
similar even when the nuclear wave packets move differently in the two
electronic potentials, as shown by $\langle R\rangle_1$ and 
$\langle R\rangle_2$ in (b).}
\label{superposition}
\end{figure}

\begin{figure}
\includegraphics[width=8cm]{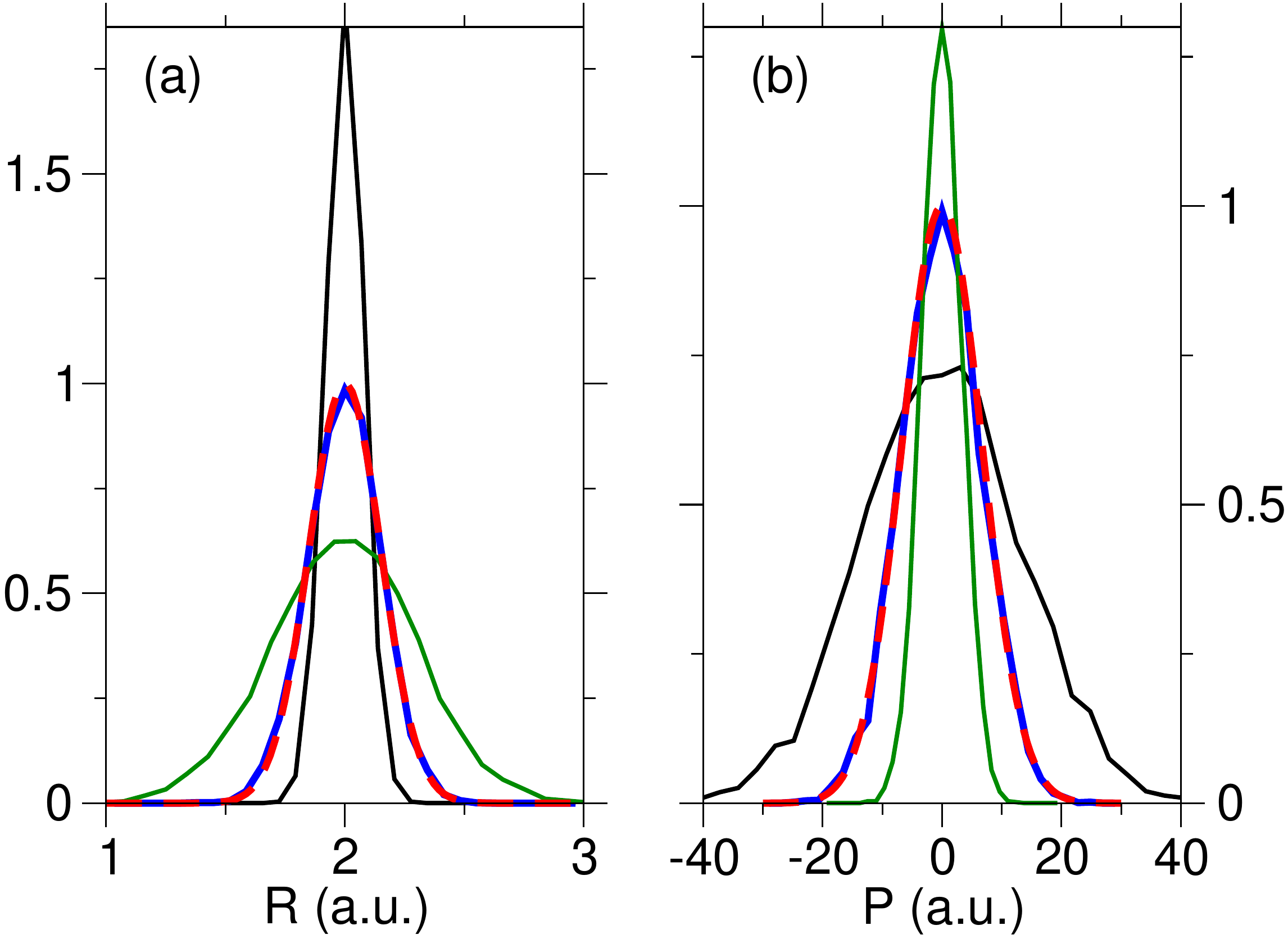}
\caption{Random sampling of $N = 10000$ initial conditions of relative
distance (a) and relative momentum (b) of the nuclei from different
Wigner distributions, the one corresponding to the quantum wave packet
(blue line) and two distributions with double or half width (green and
black lines). Also shown is dashed red line is the quantum packet in coordinate 
and momentum representations. The distributions are normalized to a maximum
value of one.}
\label{rho}
\end{figure}

In ~\ref{superposition} we compare the results of the quantum
simulation with those of the Ehrenfest approach, after averaging
$N = 10000$ trajectories. To test the sensibility of the Ehrenfest
approach to the characteristics of the nuclear ensemble of trajectories,
the initial nuclear coordinates and momentum 
are taken from random sampling of three different Wigner distributions.
One corresponds to the nuclear wavefunction of the quantum calculation 
($\tilde{\rho}(R) = \int dP \rho_W(R,P) = | \chi(R) |^2$, 
$\tilde{\rho}(P) = \int dR \rho_W(R,P) = | \chi(P) |^2$). 
The others are squeezed distributions in the coordinate or momentum
representations, as shown in \ref{rho}, as if we have doubled/halved
the width of the wave packet. 

The results of the simulation show good qualitative agreement between the 
quantum and the Ehrenfest average observables. Quantum electronic coherence 
survives several oscillations until it is completely suppressed by decoherence. 
In our model this occurs when the nuclear wave packets on the two electronic
states cease to overlap each other. 
Obviously, because we are using a one-dimensional
model and the wave packets remain bounded, revivals would be observed
at a later time.

The Ehrenfest model matches the results during the first oscillations. 
However, the classical averaging induces a faster decay than the decoherence 
rate at the beginning, although after a few femtoseconds the killing of 
coherence is also more complete in the quantum case than in the semiclassical 
approach. Regarding the nuclear motion, the average bond distance is
also very well described at short times. Although we are following the
dynamics for the first three femtoseconds, the duration is already 
significant since the bond stretches one Bohr. Of course,
in the quantum case one can separately follow the average distance in each 
electronic state, which are quite different than the mean value.

Comparing the results obtained for different Wigner distributions, 
$\rho_W(R,P)$, we find
that the induced dephasing is stronger both when we start with a wider
and a narrower distribution in the internuclear distance $\tilde{\rho}(R)$. 
The former case is expected, since we sample a wider range of initial 
positions, each originating a different trajectory, which induce the
dephasing after averaging. To understand the latter case one should
view the distribution in the momentum representation, $\tilde{\rho}(P)$, 
which by the Fourier transform is wider the narrower $\tilde{\rho}(R)$ is. 
For each system there will be an ``optimal'' initial Wigner distribution
that minimizes the dephasing rate. Just by coincidence,
in our example this distribution is approximately the one obtained
from the chosen initial nuclear Gaussian wave function.

\begin{figure}
\includegraphics[width=8cm]{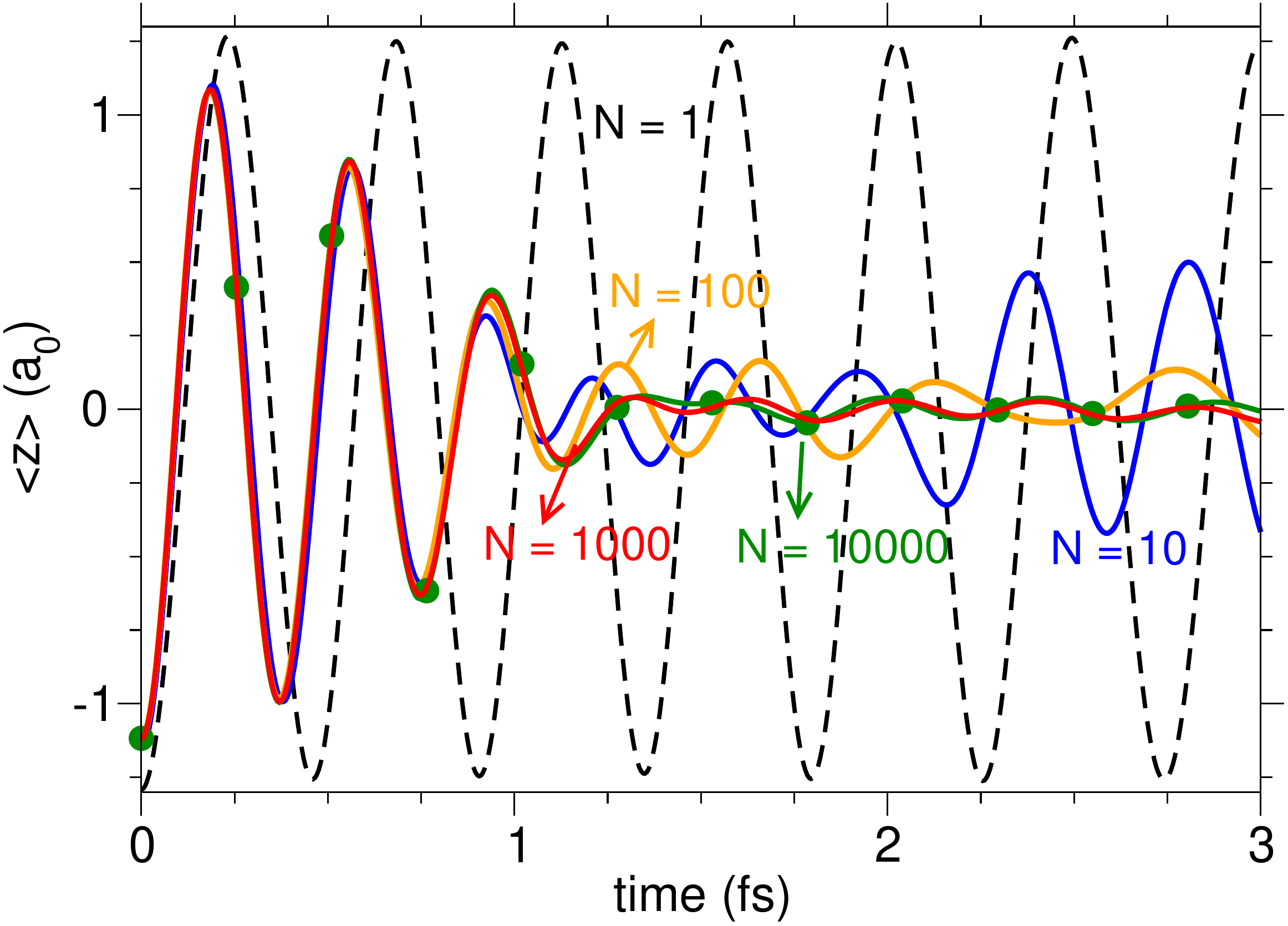}
\caption{Dynamics of the average electron position $\langle z\rangle$ obtained
from the Ehrenfest approach for different number of trajectories sampled,
$N = 1, 10, 100, 1000$ and $10000$. The results converge after $N\sim 5000$.}
\label{averages}
\end{figure}

In the semiclassical approach the results will always be sensitive 
to the number of trajectories that are included in the ensemble
average\cite{NelsonJCP12}. This is particularly important when the mean-field
departs clearly from the different electronic potentials that act on
the wave packets. Obviously, a single trajectory under this mean
field will evolve in a fully coherent way: the period of the
oscillations in $\langle z(t) \rangle$ will depend on the energy
difference between the potentials $V_2(R(t))-V_1(R(t))$, while
the amplitude of the oscillation will depend on the distance between
the turning points in the mean field potential.
In ~\ref{averages} we give an example of a typical trajectory ($N=1$) and
then show the results of the average $\langle z(t) \rangle$ as a function 
of the number of trajectories $N$ included in the ensemble.
Even a small number of trajectories can suffice to mimic the results
of the full distribution during the first oscillations. However, the
dephasing is not complete and one observes immediate revivals.
To obtain fully converged results for larger times one needs to sum over a
very large number of trajectories, which in this case is around $N \sim 5000$.

\subsection{Field-driven dynamics}


\begin{figure}
\includegraphics[width=8cm]{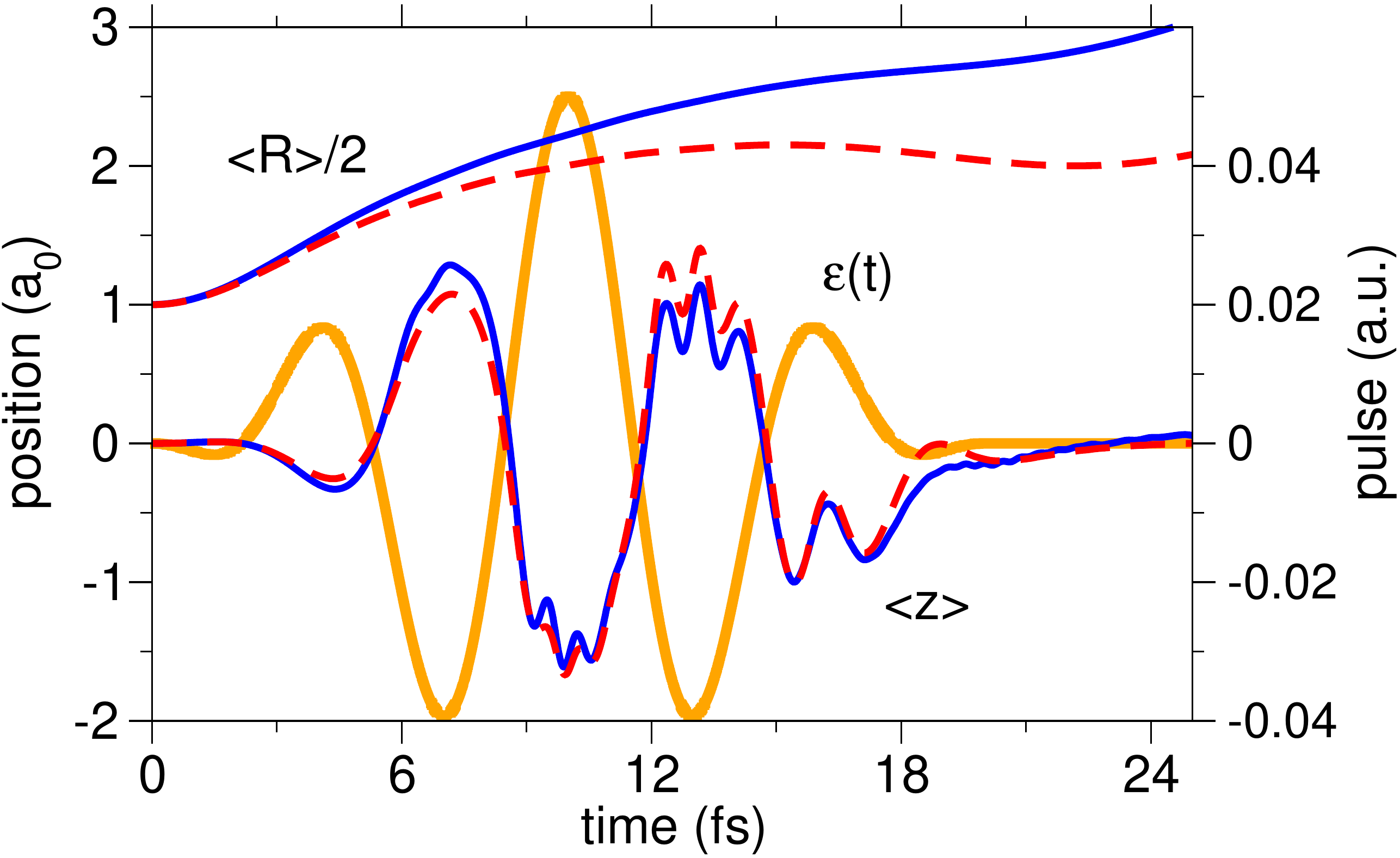}
\caption{Average electron position and bond distance after excitation
with a $3$-cycle $1.8$ $\mu$m pulse of $100$ TW/cm$^2$ peak intensity.
The dashed (red) line are the quantum results, while the continuous (blue)
line are those of the Ehrenfest model. Superimposed in orange (with scale at 
the right side of the plot) is the field amplitude showing the anti-correlation
between the dipole and the field.}
\label{shortpulse}
\end{figure}

We now consider what happens when the dynamics is driven by
a strong nonresonant laser pulse in the ground state.
The quantum results show that the 
field $E(t)$ moves the electron
(the electron position is actually anti-correlated with the field)
generating a small dipole synchronized with the external frequency
proportional to the bond distance. But the proportionality
constant is small unless the fields are very intense. On the other
hand, for too intense fields the molecule ionizes\cite{ChangJPB15b}.

In~\ref{shortpulse} we show the average electron position 
(the dipole) and bond distance using a $3$-cycle $1.8$ $\mu$m pulse of 
$100$ TW/cm$^2$ peak intensity. The Ehrenfest results match closely
the quantum dynamics for the electron, even if the deviation of the
average bond distance is noticeable.
For these pulse parameters, the Keldysh adiabatic parameter\cite{SeidemanPRL95}
$\gamma = \omega \sqrt{I_p}/E_0$ ($I_p$ is the ionization
potential from the initial state, $E_0$ the peak field amplitude
and $\omega$ the field frequency) is $\gamma \sim 0.5$, which
indicates that there is some tunneling ionization.
It is interesting to observe that Ehrenfest reproduces all the
important features of the process, including the exact form of the 
wiggles in $\langle z(t) \rangle$. Ionization in the absence of 
dissociation does not deteriorate the quality of the results.
Small deviations in the amplitudes occur. They are expected 
whenever nonadiabatic effects may be important, as induced by
the shortness of the pulse.




\begin{figure}
\includegraphics[width=8cm]{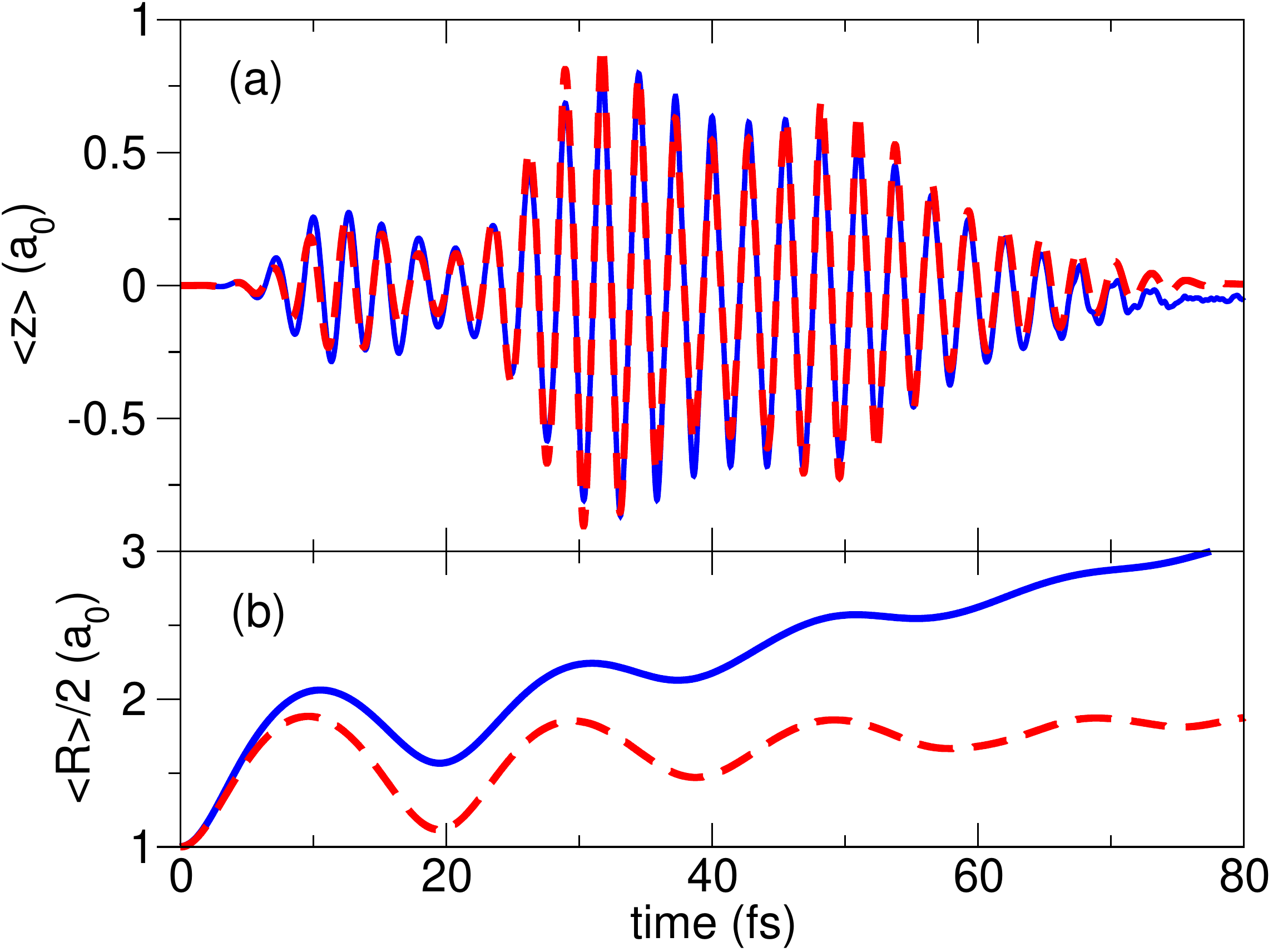}
\caption{Average electron position (a) and bond distance (b) after excitation
with a $30$-cycle $800$ nm pulse of $25$ TW/cm$^2$ peak intensity. 
The dashed (red) line are the quantum results, while the continuous (blue)
line are those of the Ehrenfest model.}
\label{longpulse}
\end{figure}

When the pulses are too short, it might be expected that the effects
of the nuclear motion on the electron dynamics are small, and hence
any failure in its description will not modify the electronic observables.
In~\ref{longpulse} we follow the dynamics driven by a $30$-cycle
$800$ nm pulse of $25$ TW/cm$^2$ peak intensity.
In this case the Keldysh parameter is $\gamma \sim 2$ and given the
pulse amplitude, multi-photon ionization is not expected (it is smaller
than 0.1\%).

Since the laser period, $2\pi/\omega$, is much shorter than the time-scale
of the molecular motion, the latter will be mostly decoupled from the electron
motion. 
As observed in \ref{longpulse} the vibrational motion in the ground
state induces a modulation on the dipole, more noticeable during the 
first oscillations which occur before the nuclear wave packet spreads.
This effect is perfectly described by the Ehrenfest model, despite the fact 
that the average bond distance stretches in time, an indication that the
mean-field potential shows bond-softening, which is not
playing an important role in the quantum dynamics.
In the quantum case, nonadiabatic effects lead to some 
dissociation (smaller than $10$\%), but the dissociating packet, after
some threshold value, is not included in the average bond distance. 
On the other hand, it always
contributes to the average potential, leading to the excess in 
bond softening simulated by the Ehrenfest approach. 
Still, the effect over the electron observables is small.



\begin{figure}
\includegraphics[width=8cm]{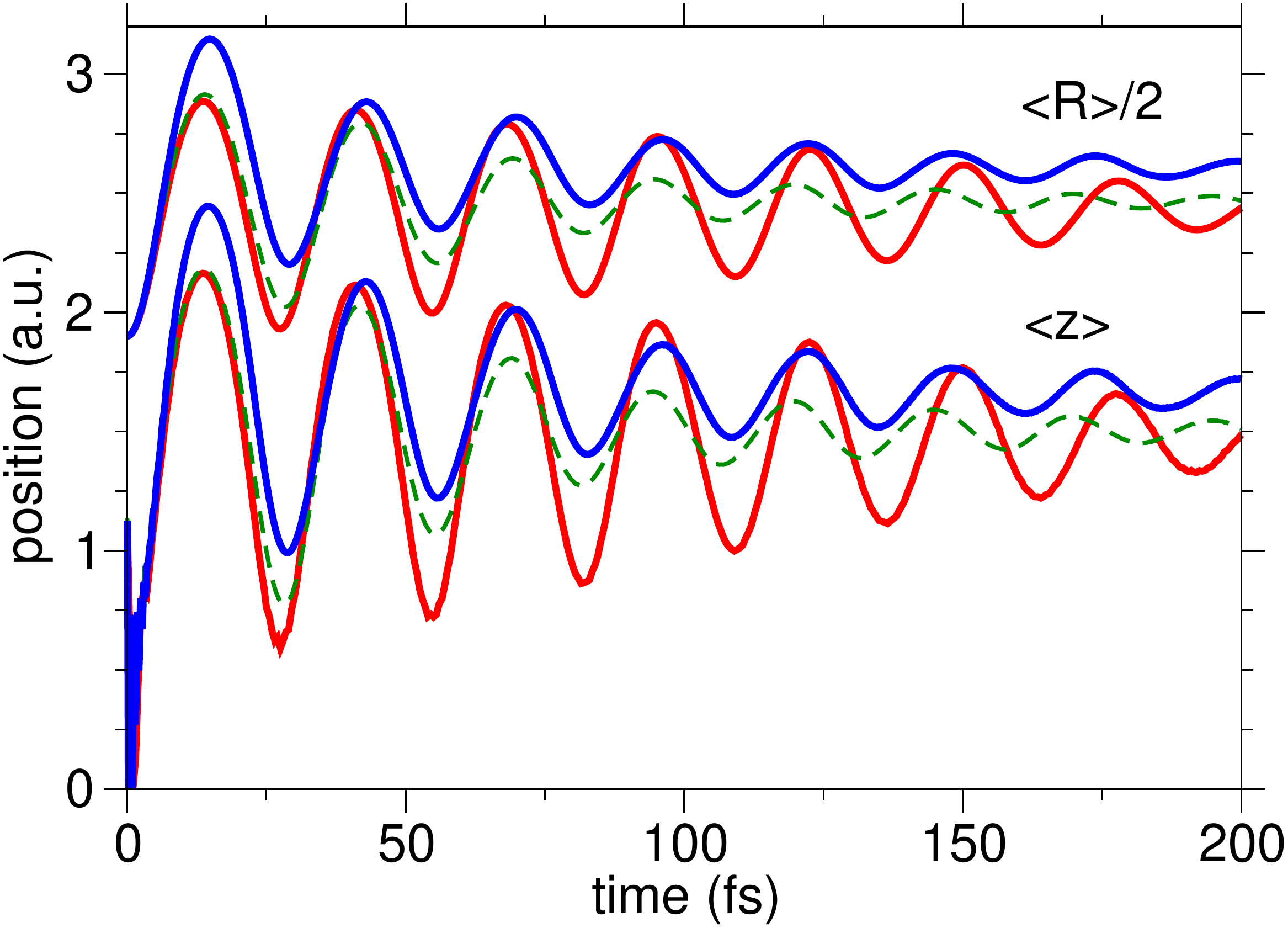}
\caption{Average electron position $\langle z \rangle$ and bond distance 
$\langle R \rangle$ for H$_2^+$ in the excited electronic state in
the presence of a strong constant field of $E_0 = 0.04$ a.u.
The red line are the quantum results and the blue the Ehrenfest results.
The green (dashed) line shows the Ehrenfest results from a wider distribution
of internuclear distances.}
\label{constant}
\end{figure}

The dynamics in the excited state in the presence of a constant field
is a fine example of fully correlated electron-nuclear motion, which
can easily be explained as the adiabatic motion of a wave packet in the
excited LIP, that shows bond hardening. In perfect conditions, for large
bond distances, the dipole is (in atomic units) 
$\langle z(t)\rangle = \langle R(t) \rangle / 2$\cite{ChangJPB15}.
The dressed electronic state implies the electron moving with the
left proton for a positive field, and the latter moving back and forth as
it reaches the classical turning points of the LIP. This creates 
the fully correlated oscillations in both observables.
As \ref{constant} shows, both the dipole and the bond oscillations
are dumped. Behind this dumping is the nuclear wave packet spreading due to
the anharmonicity of the potential, as can be easily verified looking
at revivals at later times. 
As already shown in Sec.\ref{A},
the effects of the spreading of the packet can be well accounted for
by averaging trajectories. In this example convergent results can
be obtained with $N\sim 1000$ or less.


Small nonadiabatic effects occur at early times if the initial
electronic wave function is the excited molecular electronic state, and not
the dressed state. The electron then departs from $\langle z(0) \rangle = 0$ 
and not from $\langle R(0) \rangle/2$
and some population can leak to the lowest-energy LIP where the packet
dissociates due to bond softening.
In addition, this also explains why $\langle z(t)\rangle$ is smaller
than $\langle R(t)\rangle/2$. Unlike in weaker fields, under strong fields 
like $E_0 = 0.04$ a.u.  the LIP is strongly bounded and the electron simply 
does not have enough time to overcome the initial separation and reach half
the bond distance before the nuclei get to the turning point of 
the potential and both the bond and the dipole shrink.
 
The effect of the dissociation is always reflected on a slightly weaker
mean-field potential which leads to a small relaxation in the
Ehrenfest values of $\langle R(t)\rangle$. Due to the electron-nuclear
correlation this is reflected also in $\langle z(t)\rangle$ in this
case. Interestingly, this effect can be partly compensated
for by changing the initial Wigner distribution. Sampling a wider
initial distribution of internuclear bond distances the average values
of the Ehrenfest method get closer to the quantum values, although
at the expense of having (as expected from the results of Sec.\ref{A})
over-dumped oscillations due to the larger decoherence.

\section{Conclusions}

In this work we have developed and test an algorithm that mixes grid propagation
for the electron (active) coordinates with an Ehrenfest approach to
the nuclear coordinates. We have simulated the dynamics of a collinear
model for H$_2^+$ with or without an external strong field, starting
in the electronic ground state, excited state, or a superposition of states
and compared the results with those obtained by solving the full electronic
plus nuclear TDSE of the system. The results show quantitatively good
agreement between the methods, at least for short enough times when the
dissociation is not prevalent. For sufficient sampling of trajectories,
in general we observe that the semiclassical results tend to over-characterize
the initial dephasing and the bond stretching. The first problem can be
at least partially handled by modifying the initial Wigner distribution or
biasing the sampling. The mean field results cannot reproduce the dynamics
when the wave function branches over electronic states with
very different forces, so that one cannot expect good agreement for the
long-time behavior of the nuclear dynamics in chemical reactions.
However, the results prove that the Ehrenfest approach can provide
a feasible alternative to test strong-field laser-control processes 
over transition states of molecules with several bonds, characterizing well 
ionization processes and severe bond geometrical distortions, which will
be the subject of further work.

\section*{Acknowledgments}

This work is supported by the Korean government through the Basic
Science Research program (2017R1A2B1010215) and the National
Creative Research Initiative Grant (NRF-2014R1A3A2030423),
by the Spanish government through the MINECO Project No. CTQ2015-65033-P
and by the Comunidad de Madrid through project Y2018/NMT-5028.
We thank the Army Research Laboratory for the hospitality during which
part of this work was created.

\bibliography{References}
\bibliographystyle{apsrev-nourl}

\end{document}